\documentclass[sigconf]{acmart}
\usepackage{algorithm}
\usepackage[noend]{algpseudocode}
\usepackage{tabularx} 
\newcolumntype{L}{>{\raggedright\arraybackslash}X}
\def\BibTeX{{\rm B\kern-.05em{\sc i\kern-.025em b}\kern-.08emT\kern-.1667em\lower.7ex\hbox{E}\kern-.125emX}}

\renewcommand\footnotetextcopyrightpermission[1]{} 
\begin{document}
	
	%
	\title{RevDet: Robust and Memory Efficient Event Detection and Tracking in Large News Feeds}
	
	%
	\author{Abdul Hameed Azeemi}
	\email{l154031@lhr.nu.edu.pk}
	
	\affiliation{%
		\institution{FAST-NUCES}
		\city{Lahore}
		\state{Pakistan}
	}
	
	\author{Muhammad Hamza Sohail}
	\email{l154074@lhr.nu.edu.pk}
	\affiliation{%
		\institution{FAST-NUCES}
		\city{Lahore}
		\state{Pakistan}
	}

	\author{Talha Zubair}
	\email{l154166@lhr.nu.edu.pk}
	\affiliation{%
		\institution{FAST-NUCES}
		\city{Lahore}
		\state{Pakistan}
	}
	
	\author{Muaz Maqbool}
	\email{l154053@lhr.nu.edu.pk}
	\affiliation{%
		\institution{FAST-NUCES}
		\city{Lahore}
		\state{Pakistan}
	}
	
	\author{Irfan Younas}
	\email{irfan.younas@nu.edu.pk}
	\affiliation{%
		\institution{FAST-NUCES}
		\city{Lahore}
		\state{Pakistan}
	}
	
	\author{Omair Shafiq}
	\email{omair.shafiq@carleton.ca}
	\affiliation{%
		\institution{Carleton University}
		\city{Ottawa, Ontario}
		\state{Canada}
	}
	


	%
	
	%
	
	\begin{abstract}
    With the ever-growing volume of online news feeds, event-based organization of news articles has many practical applications including better information navigation and the ability to view and analyze events as they develop. Automatically tracking the evolution of events in large news corpora still remains a challenging task, and the existing techniques for Event Detection and Tracking do not place a particular focus on tracking events in very large and constantly updating news feeds. Here, we propose a new method for robust and efficient event detection and tracking, which we call  RevDet algorithm. RevDet adopts an iterative clustering approach for tracking events.  Even though many events continue to develop for many days or even months, RevDet is able to detect and track those events while utilizing only a constant amount of space on main memory. We also devise a redundancy removal strategy which effectively eliminates duplicate news articles and substantially reduces the size of data. We construct a large, comprehensive new ground truth dataset specifically for event detection and tracking approaches by augmenting two existing datasets: w2e and GDELT. We implement RevDet algorithm and evaluate its performance on the ground truth event chains. We discover that our algorithm is able to accurately recover event chains in the ground-truth dataset. We also compare the memory efficiency of our algorithm with the standard single pass clustering approach, and demonstrate the appropriateness of our algorithm for event detection and tracking task in large news feeds. 
	\end{abstract}


	%
	\maketitle
	\pagestyle{plain}
	
	\section{Introduction}
	Internet today has become the primary source for creation and widespread dissemination of news articles leading to generation of huge amounts of news data  each day. With this unprecedented increase in the information available online, one of the major challenges is providing the user with better information navigation capability. In this scenario, automatic event-based organization of news data can lead to better structuring and classification of textual news articles data from a variety of online news media sources, and thus provide users with a better online experience. The task of automatic, event-based organization of textual news article data is named as event detection and tracking (also referred to as topic detection and tracking in many contexts). The process of discovering a new event in a stream of news articles is referred to as Event Detection. Event tracking involves the identification of further news stories that discuss the detected event, and provide some additional information indicating that the event has developed. Hence, the major task of event tracking techniques is in essence the identification of relationships between the news articles based on the event they report.

	The existing techniques for Event Detection and Tracking do not place a particular focus on tracking events in very large, complex and constantly updating  news feeds. 
			\begin{figure}[h]
			\centering
			\includegraphics[width=\linewidth]{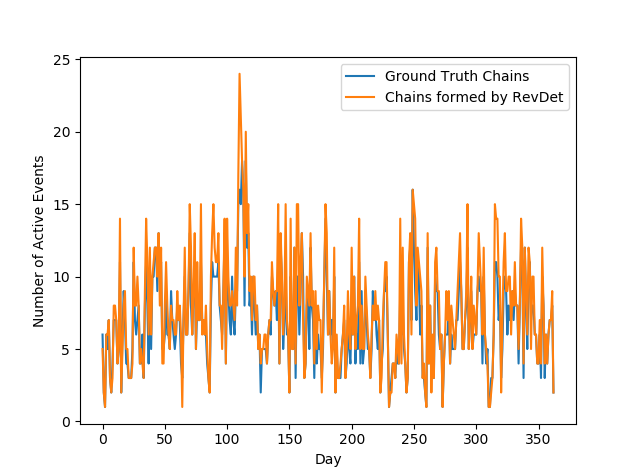}
			\caption{Per day active event chains of an year formed by our RevDet algorithm vs the ground truth chains. To form these chains, RevDet only utilized memory required for storing eight days data.  }
			\label{fig:activeeventchains}
		\end{figure}

		\begin{figure*}
		\includegraphics[width=\textwidth]{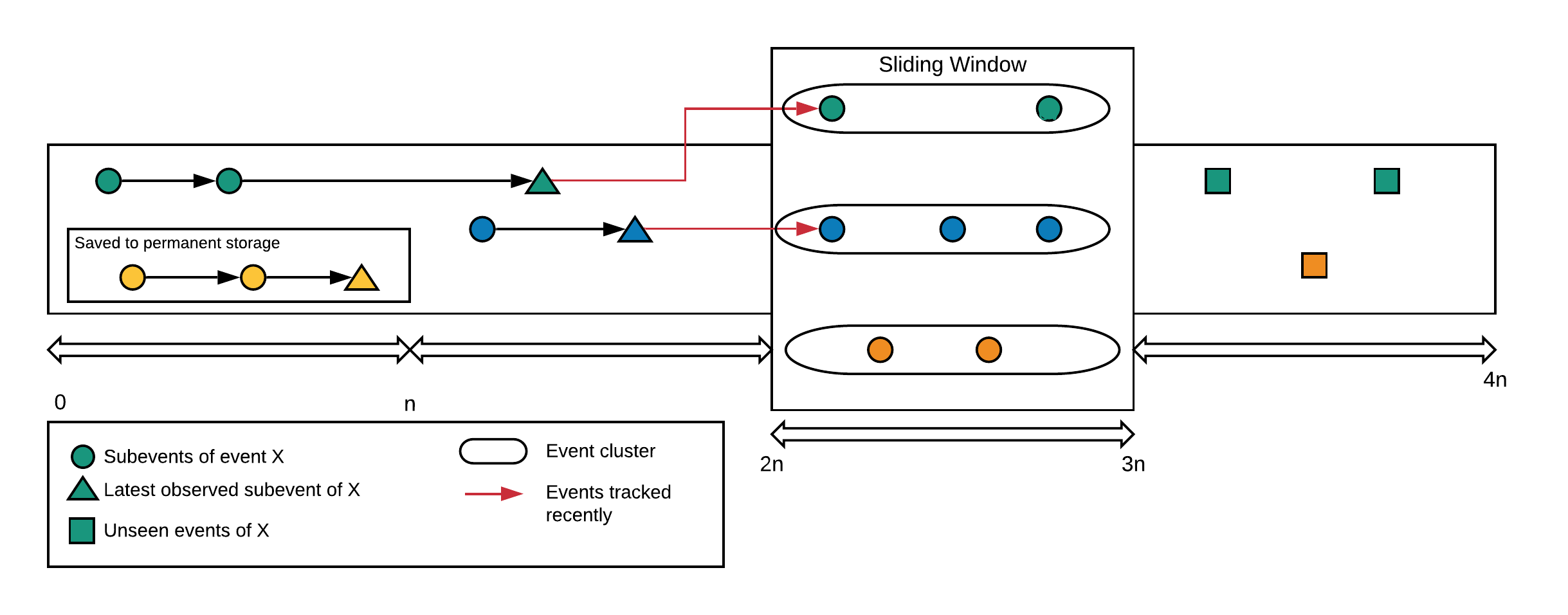}
		\caption{RevDet maintains a sliding window of size n for performing memory efficient event tracking. At any given time, only the latest observed subevents and the events in the sliding window are kept in the main memory. The events inside the window are clustered together through the Birch clustering method. After clustering, similarity between the earliest events in these clusters and the latest representatives (represented by triangles) is computed. Similar sub-events are joined together (represented by red arrow) to form an event chain. The events which are not tracked further are written to the permanent storage as a complete event chain. The event window then slides by n days and this procedure is repeated until the last event. }
		\label{fig:revdetalgorithmpic}
	    \end{figure*}
	    
The major challenge of applying Event Detection and Tracking techniques to very large news feeds is coping with the Variety, Velocity and Volume (3V's of Big Data) of such databases. Large and constantly updating news feeds exhibit the following properties:
	\begin{enumerate}
		\item Most of the events occurring across the globe are reported by multiple news agencies adding a great deal of redundancy in news feeds and a significant increase in volume. 
		\item News articles reporting a rapidly developing event tend to occur in bursts and are similar in the mention of locations \textit{i.e.} they exhibit strong spatio-temporal correlation.  
		\item Relationships between news articles are not always easy to identify from their text, with the objective details of the event being obscured by the reporting style used in the news article \textit{e.g.} a news article discussing a recently occurring event may give references to multiple events in the past, thereby complicating the extraction of correct event details from the news article.
	\end{enumerate}

	The key to developing robust and efficient approaches for event detection and tracking in large news feeds lies in taking each of these properties into special consideration. In this paper, we propose a new method for event detection and tracking, which we call the RevDet algorithm. In our method, we adopt an iterative clustering approach for tracking events by using only a constant amount of space.  Even though many events continue to develop for many days or even months, our method is able to track such events and form chains with a window-size set to a  small time unit of eight days. We also devise a redundancy removal strategy which effectively eliminates duplicate news articles and substantially reduces the size of data. Moreover, instead of utilizing all of the content of news articles, we develop a concise representation using only the article's title and a list of locations.
	For evaluating our algorithm, we also construct a large, comprehensive new ground truth dataset by augmenting two existing datasets: w2e and GDELT. We implement RevDet algorithm, and evaluate its performance on the ground truth event chains. We discover that our algorithm is able to accurately recover event chains in the ground-truth dataset, with precision of 0.82 and an  $F_{1}$ score of 0.66. We also compare the memory efficiency of our algorithm with the standard single pass clustering approach and demonstrate the appropriateness of our algorithm for event detection and tracking task in large news feeds.

	\section{Related Work}

	\begin{table*}[!htbp]
		\centering
		\begin{tabularx}{\textwidth}{|X|X|X|X|X|} 
			\hline 
			& \textbf{Segmentation} & \textbf{Retrospective Detection} &\textbf{ Online Detection}   &\textbf{Tracking} \\ 
			\hline
			CMU & Content based and lexical features for finding shift in topics & Incremental Clustering & Incremental Clustering with detection threshold and time window & K nearest neighbor and decision tree classifier \\ 
			\hline
			Dragon & Finding transitions between background topics & \multicolumn{2}{c|}{K Means clustering with modified distance measure} & Adaptation of the segmentation approach \\
			\hline
			UMass & Content based LCA (Local Context Analysis) segmentation & Bottom up agglomerative clustering & Query representation and belief threshold & Relevance feedback \\
			\hline
		\end{tabularx}
		\caption{Comparison of the approaches in TDT Initiative  }
		\label{tab:tdtinitiative}
	\end{table*}

	The task of Topic Detection and Tracking (with the topic meaning an event) was first conceived in 1996 and evolved as a joint venture between University of Massachusetts, Carnegie Mellon University and Dragon Systems \cite{allan2003topic}. It was a yearlong pilot study focusing on segmentation of data streams, identifying events in news stream and tracking a particular event in different news. This initiative provided grounds for further research on this topic and established some initial techniques and methodologies to address the problem. 
	
	The problem was divided into three main tasks: 
	\begin{enumerate}
		\item Segmentation of the data/news stream into distinct, topically homogenous blocks.
		\item Identification the first occurrence of a news story discussing a new event.
		\item Subsequent tracking of the news stories that discuss the event.
	\end{enumerate}
	
	Three approaches were used for the task of segmentation, detection and tracking: Dragon's Approach, UMass Approach and CMU Approach. Dragon's approach for segmentation treated a stream of data as a sequence of unlabeled topics. The task of segmentation was then reduced to finding the transitions between these topics.  These topics in the first place were constructed using a multi-pass k means algorithm which grouped the news stories into k clusters, with each cluster corresponding to a distinct topic. These clusters were used to build background language models, which were then used for the segmentation task. All the three approaches are summarized in Table \ref{tab:tdtinitiative}.

	Existing event detection and tracking algorithms usually adapt the single pass clustering algorithm for the identification of news events \cite{hasan2019real, osborne2014real, becker2011beyond} . For each incoming news article, its similarity with previous known events is computed. If the similarity exceeds a similarity threshold, the news article is flagged as referring to an existing event. Otherwise, the news article is classified to be a new event. The inherent problem in the application of single pass clustering algorithm to very large event data is evident: the single pass clustering algorithm must maintain a "memory" of news events. Although this is feasible for small datasets such as TREC, maintaining all the events in memory quickly becomes a significant challenge if large news feeds are dealt with, due to the scale at which events are reported each day all over the world. Alternate forms of news representation such as forming a query with only named entities and quantitative details fails to address the problem; important details form a significant part of news articles, and rigorous preprocessing for a significant reduction in size of news article in memory can lead to misclassification and a significant drop in precision. 
	
	The solution to this problem is to extend the concept of a growing 'entropy' of the news article as used by Radinsky \cite{radinsky2013mining} i.e. penalizing
	an event on the time distance between two events. Experiments on the TDT4 corpus with different time thresholds have shown n=14	days threshold to be the most appropriate. If we make this a binary threshold, we will need to place only n days data in the memory, with n being the upper limit on the number of days between any	two events as determined by experiments.
	
	Other systems have considered tracking more generic 'topics' within the news articles. In \cite{leskovec2009meme} a framework has been presented for tracking topics in news articles via short, distinct phrases that remain intact throughout the articles. Our focus, however, is to develop a technique for tracking 'events' instead of 'topics' in news stories, which are more specific and require a much greater context than just a few phrases for achieving a high precision. Some efforts have been made to leverage topic modelling for detection and tracking of news events. One such technique, Latent Dirichlet allocation (LDA), is widely used in detecting events through posts on micro-blogging sites such as Twitter. Diao et al  \cite{diao2012finding} developed an LDA model which is able to find bursty topics on Twitter by capturing two phenomena: posts by same user or around same times are more likely to correspond to same topic/event.  Hierarchical Dirichlet Process (HDP) is used by Srijith et al \cite{srijith2017sub} for detecting sub-stories/ sub-events by learning subtle variations in them through topics underlying events. A variation of HDP incorporating time dependency, paired with mixture Gaussian model is used by Wang et al \cite{wang2013real} for detecting bursty words and newsworthy events in Twitter data. LDA with selectional preferences (LDA-SP) has been used in combination with ConceptNet by Vo et al \cite{vo2015exploiting} to detect events by first identifying relationships between entities mentioned in Twitter posts, and then classifying them using k-nearest neighbours technique. Huang et al \cite{huang2012microblog} showed that using LDA together with the single pass clustering algorithm to deal with the short and sparse microblog data decreases miss and false alarm rates.  
	
	Graph-based modeling approaches have also been used in Event Detection and Tracking. Sayyadi et al \cite{sayyadi2009event} presented an approach of building a KeyGraph ( graph containing extracted terms \cite{ohsawa1998keygraph}) with keywords having lower inverse document frequency (IDF) filtered out. Connected key words are those that occur in same document. Closely related words form a community. A community is considered to be a synthetic document and titled as a key document. Clustering then groups together documents similar to the key documents, and each cluster is considered as an event.
	
	Many event detection and tracking algorithms tend to perform better on carefully curated test datasets but struggle to generalise to real world news feeds. To the best of our knowledge, this is the first attempt to devise an event detection and tracking strategy for large, noisy and complex news feeds containing a great portion of duplicate news articles.

		\begin{figure*}
		\includegraphics[width=\textwidth]{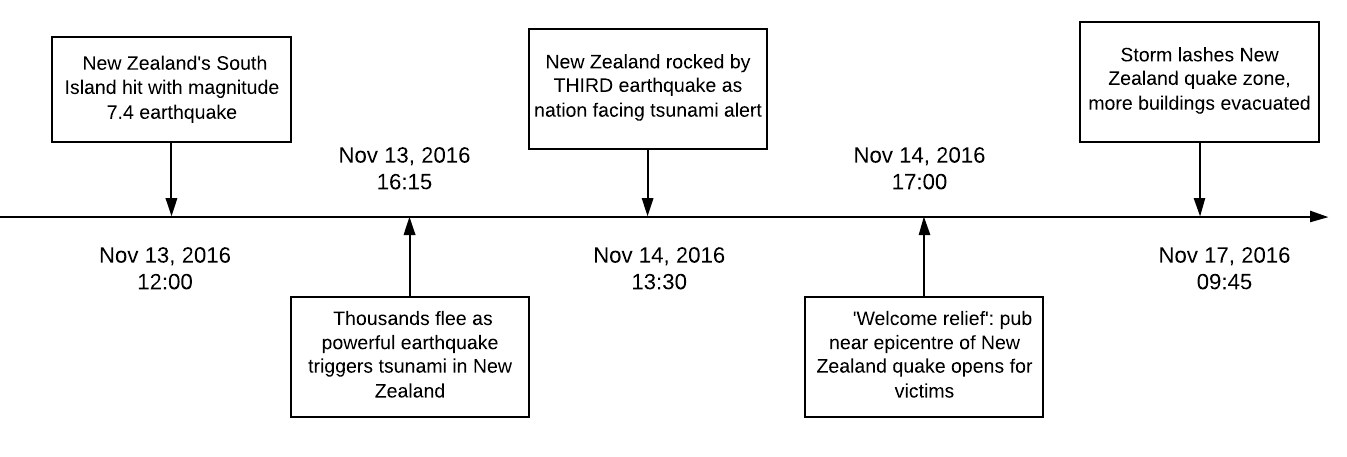}
		\caption{An event chain showing the progress of subevents related to earthquake in New Zealand.}
		\label{fig:eventChainExample}
	\end{figure*}
	
		\section{Definitions}
	{\bf Event} An event is an occurrence at a particular location during a particular interval of time. An event is further composed of {\bf subevents} such that the beginning and the end of an event correspond to two separate subevents. Since we are dealing with online news data, we will consider newsworthy events only \textit{i.e.} events that are significant enough to be reported by at least one online news agency.   \\
	{\bf Subevent} A subevent, which is an atomic part of an event, is an occurrence at a particular location and time. A subevent may only be a part of one event only.  \\
	{\bf News Article} A news article $a$ represents a subevent \textit{e} is characterized by its publication timestamp $t$, title $h$ and a list of locations mentioned in the article $l$. 
	\begin{equation}
	a = (t, h, l)
	\end{equation}
	{\bf Event Chain} An event chain $C$ is an ordered set of sub events  $ \{e_{1}, e_{2}, e_{3}, . . . , e_{n}\} $ of a particular event , sorted in increasing order of timestamp and where each new sub event has some additional information as compared to its predecessor.
	\begin{equation}
	C = \{a_{1}, a_{2}, a_{3}, . . . , a_{n}\}
	\end{equation}
	{\bf Latest Representative (LR)} Latest representative of an event $a_{1}$ is the sub-event in event chain with the latest timestamp \textit{i.e.} the most recent news about an event.\\
	{\bf Earliest Representative (LR)} Earliest representative of an event ${a_n}$ is the sub-event in event chain with the smallest timestamp \textit{i.e.} the first news about an event.\\
	{\bf Event Window} 	Event window consists of unordered subevents of different events occurring in a particular time frame $\Delta t$.
	
	\section{Approach}
	
	The first step in devising an approach for event tracking is to consider what makes an event different from others. Depending on this definition of an event, the event chains formed may be considerably different \textit{e.g.} an event chain of a general election in a certain country may involve all the news in relation to the election or only the news relating to the rallies by one candidate. The decision of this is made by determining what constitutes the event identity \cite{allan1998line} , which is something unique to every new event, and common to the sub-events in event chain.  If an event is taken to be something that happens at particular place and time, then the locations mentioned in a news story and $t \pm n  $ days  constitutes the identity of event, with $t$ being the event timestamp. Another option could be to include named entities \textit{e.g.} \textit{people, organizations} as part of event identity, and this has been seen to considerably increase recall in event tracking tasks \cite{radinsky2013mining}.
	
	{\bf Selecting the clustering algorithm} 
	The choice of clustering algorithm for the formation of event chains is an important one, since it directly influences the representation of news articles, quality of chains formed and efficiency of the approach. Some approaches have used the k nearest neighbours algorithm for finding closest news articles or the k-means algorithm for grouping together the related news. These methods would fail to work in a big data setting since they require a parameter $k$ as input \textit{i.e.} the number of articles to group together. Other algorithms include Wave-Cluster, DBSCAN and BIRCH. Wave-Cluster is a grid based algorithm and the main advantage of this algorithm is the fast processing time \cite{6832486} . However, Wave-Cluster does not perform well for our problem as using a single uniform grid does not result in good quality clusters nor does it satisfy the time constraints for a highly irregular data distribution (news articles). DBSCAN is a density based clustering algorithm. It can efficiently deal with noise while forming high quality clusters. Unlike k-means, DBSCAN does not require the input parameter k which is used to identify the number of clusters to be formed. Although, DBSCAN seems to be a good choice, the major drawback of this algorithm is its inability to efficiently cluster data sets with large differences in densities. 
	
	BIRCH \cite{zhang1996birch} is an unsupervised hierarchical clustering algorithm suitable to cluster large data sets. The main advantage of using BIRCH is that it can work incrementally \textit{i.e.} does not require the whole data set in advance and can efficiently adjust the number of clusters to be formed relative to the input data set. BIRCH typically requires a single scan of the data set to form good quality clusters and this quality can be improved using additional scans if required.
	
	Similar to DBSCAN, BIRCH can work without the input parameter k and can decide for itself the number of clusters to be formed. This feature of BIRCH is essential for our research problem as the number of event chains present in a given set of news articles can vary. Moreover, BIRCH is a first of its kind algorithm that can efficiently handle noise. News articles which do not progress are considered noise in our case as they form an event chain consisting of only one node i.e. they are not tracked further.
	
	{\bf Representation of news articles} .
	We represent every news article as a vector of title, themes, locations and counts contained within the news article. \\
	\begin{enumerate}
		\item \textbf{Title} of the news article reporting the event. \\ \textit{Example:}  Powerful earthquake strikes New Zealand killing 2 people.
		\item \textbf{Themes} associated with the event. \\ \textit{Example:} NATURAL DISASTER; NATURAL DISASTER EARTHQUAKE; CAUTION ADVICE; KILL; 
		\item \textbf{Locations} contained within the news article.  \\ \textit{Example:}  Wellington, New Zealand, (Lat, Lng): -41.3,174.783
		\item \textbf{Counts} associated with the event reported by the article, and of a particular location.  \\ \textit{Example:} KILL 2, New Zealand, NZ;
	\end{enumerate} 
	 These fields have been pre-extracted for every article in GDELT GKG. Instead of a \textit{tf-idf} representation, we convert themes, locations and counts into one-hot vectors, and use a sparse representation of these vectors. This type of representation is readily accepted by the existing implementations of the Birch algorithm \textit{e.g.} SciKit \cite{scikit-learn} implementation of Birch.  
	
\begin{figure}[h]
		\centering
		\includegraphics[width=\linewidth]{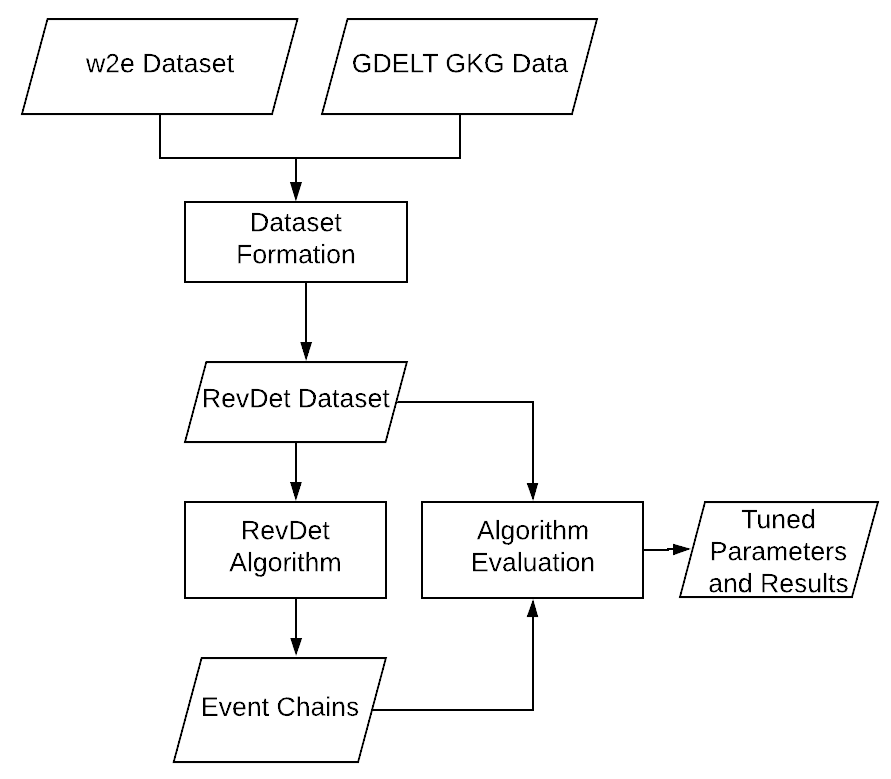}
		\caption{A high level overview of the approach taken for formation of event chains and evaluation of results }
		\label{fig:workflow}
	\end{figure}

	 Figure  ~\ref{fig:workflow} gives an overview of the workflow adopted for forming event chains on the prepared dataset through the proposed algorithm, and evaluating the results.
	
	\section{RevDet Algorithm}

	Now, we describe our algorithm to form event chains from news data. We say that every sub event $x$ in an event chain $C_i$ contains sufficient information that enables tracking of further events solely through $x$, and that these subevents cannot track events of some other event chain $C_j$ . i.e. for every $a$, $b$ in an event chain $C_i$,
	        \begin{displaymath} similarity(C_a^i, C_b^i) > \theta > similarity(C_a^i, C_x^j) \mid i \neq j \end{displaymath}
	In other words, if we are presented only with the first event of a chain, we will be able to recover the whole event chain from the news feed. We adopt an iterative clustering approach for tracking events. 
	\begin{enumerate}
	    \item Initially we add first $n$ days data to the event window.
	    \item Then we cluster articles data through the birch clustering algorithm, and save the resultant event chains to permanent storage.
	    \item Now, we extract the latest representatives of these event chains, and keep them in temporary storage, discarding the rest of data in the chains (at any given time, we only keep latest representatives belonging to at most one event window in the past). We slide the event window by $n$ days. 
	    \item  We then again cluster articles data to form chains $y$. 	For each latest representative $l_i$ saved in the previous step , we compute its similarity with each of the earliest representative $e$ of $y$. 	\begin{displaymath} sim(l_i, e) \end{displaymath} where the similarity of two events $sim(a, b)$ is defined as:
		\begin{displaymath} 
	 jaccard(a_{title}, b_{title}) * jaccard(a_{location}, b_{location})	\end{displaymath}  	This ensures that two events are be considered similar if they both belong to a certain subject (represented by title) and occur in proximity (represented by location).

	    \item If the similarity is greater than 0, this indicates that the event has developed; hence we merge these event chains with their previous one. Otherwise, we save the event chains $y$.
	\end{enumerate}
	
  This whole process is repeated until event window reaches the end. The overall RevDet method is outlined in Algorithm (1).

	\begin{algorithm}
		
		\caption{Event Chain Formation}\label{euclid}
		\begin{algorithmic}[1]
			\Procedure{RevDet}{$days$,$windowSize$,$threshold$}
			\State $fileIndex \gets dict()$
			\State $fIndex \gets 0$
			\State $latestRepresentatives\gets []$
			\State $i\gets 0$
			\While{$i\le n$}
			\State $previousWindow \gets days[i-windowSize: i]$
			\State $latestRepresentatives.keep(previousWindow)$
			\State $end\gets i + windowSize$
			\State $data\gets getData(days[i:end])$
			\State $df\gets concat(data['title'], data['locations'])$
			\State $df\gets oneHotEncode(df, sparseOutput = true)$
			\State $clusters\gets birchClustering(df, threshold)$
			\For{cluster in clusters}
			\State$filePath\gets str(fIndex)+ ".csv"$
			\State$fIndex++$
			\State $eR\gets getEarliestRepresentative(cluster)$
			
			\For{row in latestRepresentatives}
			\State $s1\gets jaccardSimilarity(row.title, eR.title)$
			\State $s2\gets jaccardSimilarity(row.location, eR.location)$
			\If {$s1 > 0 $ and $s2 > 0$}
			\State $filePath\gets fileIndex[eR.id]$
			\State $connectedEvents\gets getData(filePath)$
			\State $df\gets concat(connectedEvents,df)$
			\State $latestRepresentatives.remove(row)$
			\EndIf
			\EndFor
			\State $lR\gets df.tail()$
			\State $fileIndex[lR.id]\gets filePath$
			\State $latestRepresentatives.concat(lR)$
			\State $df.sort()$
			\State $df.writeToFile(filePath)$

			\EndFor
			\State $i\gets i+windowSize$
			\EndWhile
			\EndProcedure
		\end{algorithmic}
	\end{algorithm}

	\subsection{Implementation}
	We have implemented the RevDet algorithm \footnote{Code and Data are available at \url{https://github.com/ahazeemi/RevDet} } in Python on top of the Birch Clustering Algorithm available in SciKit Learn \cite{scikit-learn}. Our algorithm takes as input news articles data (with two necessary columns: a list of locations and title) in the form of per day files (sorted by ascending timestamp of the event), window size and birch threshold. It then forms event chains and outputs each chain in a separate file. During the formation process, it also writes some temporary files to the permanent storage, and removes them once all chains have been formed.
	
	\section{Experiments}
	
	\subsection{Dataset}

	\begin{figure*}
		\includegraphics[width=\textwidth]{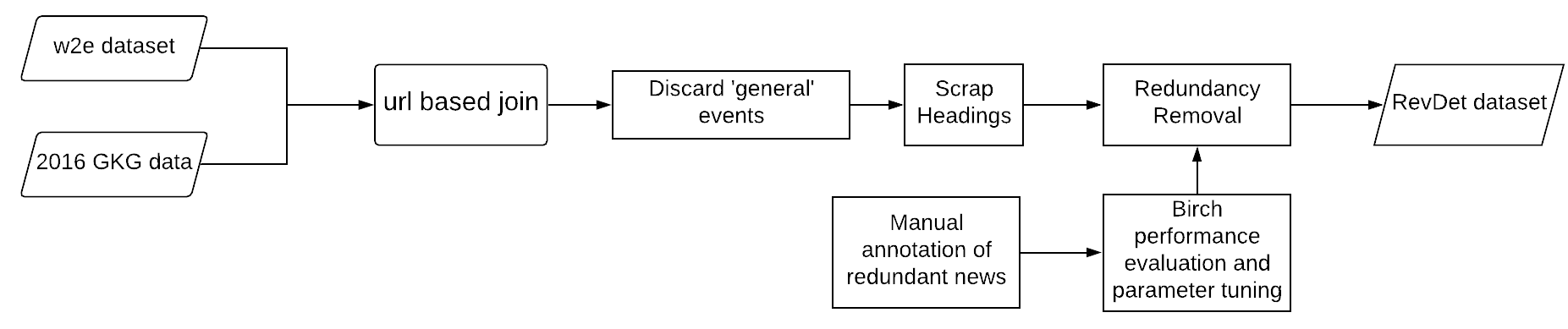}
		\caption{Formation of the RevDet dataset for evaluating event detection and tracking approaches.}
		\label{fig:datasetFormation}
	\end{figure*}

	{\bf GDELT}
	
	GDELT \cite{leetaru2013gdelt} is a real-time database of global human society, and essentially contains a large amount of processed world news .  The GDELT global knowledge graph (GKG) is a part of GDELT database, and is the largest publicly available dataset of news events across the globe. It contains processed data from real-time news from around the world including locations, themes, organizations, people and tone of every news event. The GKG table in the GDELT database has 27 columns containing a wealth of information about each news article. This dataset provides us with pre-extracted fields of each news article for running our Event Detection and Tracking algorithm. 
	
	Along with this, we require a fairly large event tracking dataset with fine-grained ground truth for an effective evaluation of our algorithm. TREC's TDT's datasets are unsuitable for this purpose, as they are obsolete and small: they were collected in the year 2000 and have around only 13k articles grouped in 279 topics.  The recently released dataset w2e \cite{hoang2018w2e} is a manually constructed substantially large TDT dataset containing 207,722 events grouped in 4501 events and 2015 event chains.  Each event chain contains urls of news articles and short text describing each subevent in the chain. 
	
	Although w2e contains a short description of each event, it lacks the specific processed details of news events as available in GKG (themes, locations, tone etc.). To address this problem, we reconstruct the w2e dataset by augmenting it with the GDELT dataset \textit{i.e.} each url in the original w2e dataset is searched in GDELT GKG table, and the details contained in the matched row in GKG table are appended to w2e. From the resultant data, we keep only the chains which adhere to the concept of event defined earlier \textit{i.e.} throughout its development, a news event must contain similar locations. This process discards chains with a more general topic for example a chain containing all news related to the US Presidential Election, instead of a specific event. Following this process (Figure ~\ref{fig:datasetFormation}), we are able to construct a fairly large and a rich dataset: RevDet dataset, for evaluation of our event tracking algorithm containing 1329 event chains.

	\subsection{Redundancy Removal}
	Most of the events are cited by multiple news agencies across the globe, thereby adding a substantial amount of redundancy to data in news feeds. This redundancy needs to eliminated since two news articles referring to the same subevent would occur as two nodes in an event chain, with the latter node providing no upgraded knowledge about the event. For removing this type of redundancy in news articles, we utilize the birch clustering algorithm for clustering news articles on various attributes like Themes, Locations and Counts. Now, we have four different methods for performing clustering on these attributes:
	
	\begin{enumerate}
		\item Clustering on title and locations first, then sub-clustering the resulting clusters on the basis of counts,
		\item Clustering on title, then sub-clustering on locations and counts, 
		\item Clustering on title, locations and counts, or
		\item Clustering on locations, then sub-clustering on title and counts
	\end{enumerate}

	To compare the performance of these four methods and tune birch parameters, we manually cluster a subset of GKG data of 354 news articles containing 7 events and construct a ground-truth dataset containing clusters of duplicate news articles. Two news are grouped together only if they represent the exact same subevent. It is important to note here that while they contain the same information, they are two different news articles with possibly different reporting styles and the choice of words. Hence, our task is tailored towards news data and slightly different from the approaches for near-duplicate detection, which are more general and do not consider specific properties of news articles like title, locations and counts etc. 
	
	We evaluate performance by clustering the news articles and comparing to the ground truth clusters. Clustering accuracy is evaluated by calculating Precision, Recall and  $F_{1}$-Score over pairs of articles \textit{i.e.} through the pair-counting method. The precision is calculated as \begin{displaymath}P = \frac{TP}{TP+FP}\end{displaymath} \textit{i.e.} the fraction of pairs correctly put in one cluster, and recall as \begin{displaymath}R = \frac{TP}{TP+FN}\end{displaymath} \textit{i.e.} how many actual pairs were identified.  $F_{1}$-score is the harmonic mean of precision and recall and is used for selecting the best birch parameters for each clustering approach, and we use this score for comparing the four clustering approaches (Table \ref{tab:redundantnewsclustering}). As shown, clustering on title and locations first, then subclustering on counts yields the best result making it a suitable approach for removing duplicate news articles. This procedure results in a 57\% decrease in the data size. \\
	
	\begin{table}

		\scalebox{0.9}{
			\begin{tabular}{lllll}
				\hline
				First Level              & Second Level     & Precision & Recall &  $F_{1}$ \\
				\hline
				Title, Locations         & Counts           & 0.97       & 0.77    & 0.86      \\
				Title					 & Locations, Counts& 0.75       & 0.79    & 0.77      \\
				Title, Locations, Counts & -                & 0.67       & 0.68    & 0.67      \\
				Locations				 & Title, Counts   & 0.92       & 0.41    & 0.57    \\
				
				\hline     
		\end{tabular}}
		\\
		\caption{Precision, recall and $F_{1}$ score for four different approaches of clustering redundant news. Clustering on title and locations first, and then sub-clustering on counts yields the best result, implying that the title and locations combined have the greatest discriminatory power of correctly separating two different news.}
		\label{tab:redundantnewsclustering}
	\end{table}
	
	{\bf Article's Title vs Content}
	We now consider using article's content instead of title for detecting duplicate news articles to see whether there is a significant gain in the  $F_{1}$-score. For this task, we use the themes field (originally contained in GDELT GKG) in the dataset, which describes all the themes contained in a news article through special categories and taxonomies which accurately capture the content \textit{e.g.} a news article about the destruction of roads by heavy rain contains themes like
	\begin{itemize}
		\item NATURAL\textunderscore DISASTER\textunderscore MONSOON
		\item INFRASTRUCTURE\textunderscore BAD\textunderscore ROADS
	\end{itemize} We compare the performance by first clustering duplicate news on title, locations and then counts. We repeat the same process with themes instead of title. As the results in Table \ref{tab:titlevscontent} show, using article's content (themes)  does not lead to a significant change in the  $F_{1}$ score. This shows that a news article's title has the ability to accurately and succinctly describe the event reported in it. Moreover, as the average content length of a news article in the data (represented by themes length) is significantly greater than the article's title, clustering on title is a more suitable option of removing duplicate news than clustering on article's content. 
	
	\begin{table}

		\begin{tabular}{lllll}
			\hline
			Method         & Av. length in characters & Precision  & Recall  &  $F_{1}$ score \\
			\hline
			Title         	& 18.7 & 0.97       & 0.77    & 0.86      \\
			Themes          & 16476.0 & 0.96       & 0.81    & 0.88      \\
			\hline     
		\end{tabular}
		\caption{Comparing the performance of title and content (represented by themes) for clustering duplicate news together. }
		\label{tab:titlevscontent}
	\end{table}

	\subsection{Algorithm Evaluation}
\begin{figure}[h]
	\centering
	\includegraphics[width=0.7\linewidth]{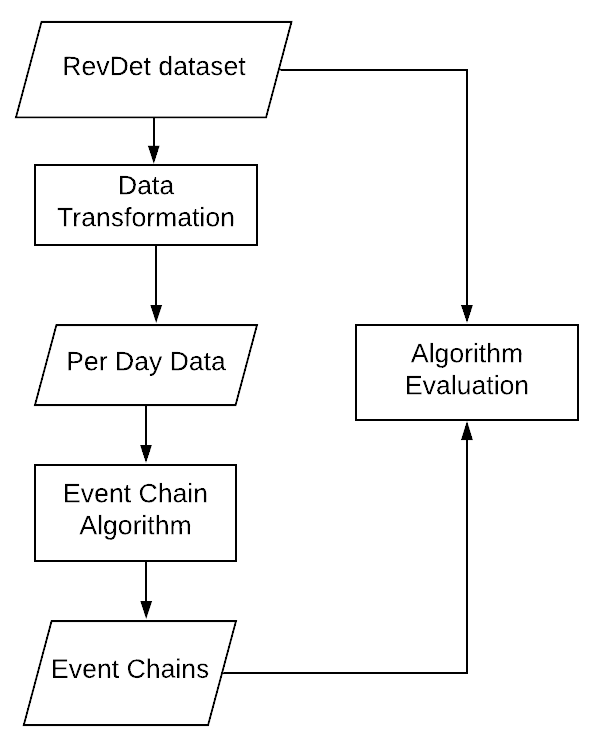}
	\caption{An overview of the steps involved in preparing data for evaluation of event chains formed by RevDet}
	\label{fig:revdetEvaluation}
	\end{figure}

	We evaluate the performance of the algorithm by comparing the event chains in the ground-truth dataset with the event chains formed by the algorithm. For this, we first transform the dataset into per day files, simulating the way in which data would be available to the algorithm in a news feed (Figure  ~\ref{fig:revdetEvaluation}). We then run RevDet on these per day files and evaluate performance through Precision, Recall and  $F_{1}$ score over pairs of articles in the ground truth event clusters and the formed clusters. 
	
	{\bf Clustering Performance}
	The best performance  of RevDet is reached on Birch Threshold 2.3, and Window Size (Table \ref{tab:results}).  The $F_{1}$ score of 0.66 on 0.82 precision is adequate enough to form event chains of good quality as have focused on precision focused tuning to avoid distortion of event chains with irrelevant news. A relatively low recall indicates the difficulty in clustering news events together with different wordings of the title. This problem can be alleviated in future work through the use of pretrained paragraph level embeddings like Doc2Vec \cite{Le:2014:DRS:3044805.3045025}. The in-memory clustering approach has a much lower precision and recall. This is due to a greater chance of an event landing in the wrong chain as the time-dependancies between events are ignored by loading all the data into the memory at start and then performing clustering. \\  
	We next present a macro-level comparison of active event chains in the original dataset and in the formed ones in Figure ~\ref{fig:activeeventchains}, which shows the number of event chains that are still being developed on each day. It can be seen that RevDet has been able to closely replicate the ground truth events. 
	
	\begin{table}

	\scalebox{0.9}{
	
	\begin{tabular}{llllll}
		\hline
		Algorithm & Birch Threshold & Window Size & Precision  & Recall &  $F_{1}$ \\
		\hline
	 	RevDet & 2.3 & 8      & 0.81    & 0.56    & 0.66   \\
	 	In-Memory & 2.2 & -      & 0.56    & 0.24    & 0.34   \\
		\hline     
	\end{tabular} }
	\caption{RevDet vs In-Memory clustering performance on tuned parameters as evaluated on ground truth chains. RevDet performs far better than the in-memory clustering approach. }
	\label{tab:results}
	\end{table}

\begin{figure}[h]
	\centering
	\includegraphics[width=\linewidth]{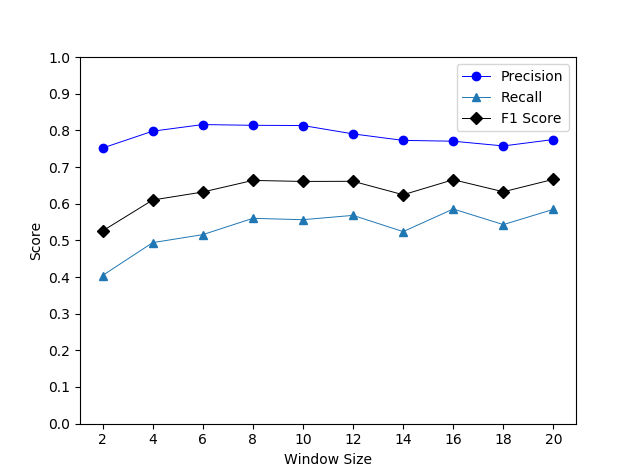}
	\caption{Plot of precision, recall and  $F_{1}$ score vs. window size of RevDet. At window size 8, RevDet is able to track events with almost same clustering accuracy as with window sizes closer to 20, while needing much lesser memory.}
	\label{fig:algorithmResults}
\end{figure}
	
	{\bf Window Size}
	We next focus on the effect of window size on the results (Figure  ~\ref{fig:algorithmResults}). We discover that varying the window size after 8 has little effect on the $F_{1}$ score \textit{\textit{i.e.}} it stays between 0.64 and 0.66. This makes 8 a good choice for window size, and implies that most of event chains do not have a gap of greater than 8 days between any two consecutive news.   We also observe that the precision drops slightly as the window size is increased, owing to the greater data in the event window.

\begin{figure}[h]
	\centering
	\includegraphics[width=\linewidth]{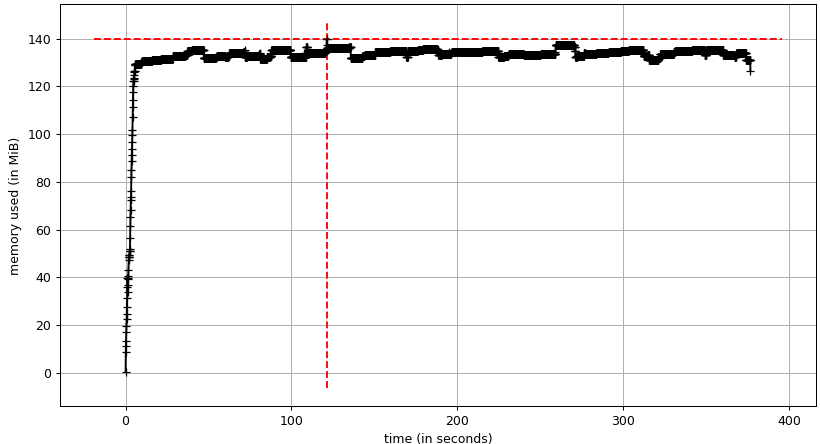}
	\caption{Memory usage vs running time of RevDet algorithm. The small spikes represent the movement of event chains to and from the main memory according to their development. }
	\label{fig:revdetMemoryUsage}
\end{figure}

\begin{figure}[h]
	\centering
	\includegraphics[width=\linewidth]{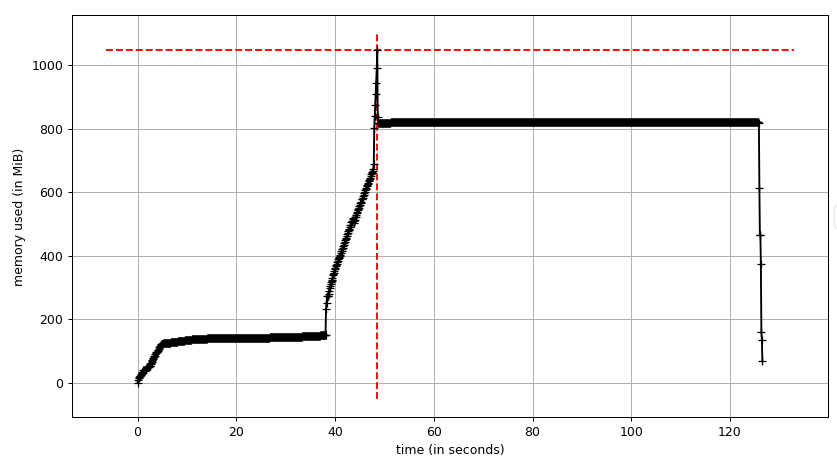}
	\caption{Memory usage vs running time of an in-memory clustering approach which loads all the data into the memory once and then performs clustering. The memory increase from 40s to 50s represents the transfer of all data to the memory; the spike at 50s is due to clustering all data through Birch at once.  }
	\label{fig:inMemoryClusteringPlot}
\end{figure}

	\subsection{Scalability of RevDet}
	We next examine the memory efficiency and scalability of RevDet. The plot in Figure ~\ref{fig:revdetMemoryUsage} shows the memory usage as the algorithm progresses. As expected, the space requirement of temporary storage (RAM) is constant with respect to the input data. The spikes are representative of the movement of event chains to and from the memory. RevDet has the ability to scale efficiently with respect to the number of news articles in the dataset which makes it a very suitable approach for event detection and tracking in large news feeds. We also examine the space requirement of the in-memory clustering approach in Figure \ref{fig:inMemoryClusteringPlot}. The memory usage rises sharply as all of news data is loaded into the main memory at the start, and becomes constant once formed chains are being written to the permanent storage. Peak memory usage of the in-memory clustering approach ($\approx 1000$ MB) is 7 times the peak memory usage of RevDet ($\approx 140$ MB).  Moreover, as the input data will increase, the memory requirements of the former approach will grow proportionally making it infeasible to form event chains. 
	
	\section{Conclusion and Future Work}
	In this paper, we have tackled the problem of robust and efficient detection and tracking of news events in large news feeds. An iterative clustering based algorithm has been proposed for this purpose which is able to extract event chains of events that continue to develop for a long period of time, using memory as low as required for clustering eight day news. We also propose a redundancy removal strategy for removing duplicate news articles. We construct a new, comprehensive ground truth dataset by augmenting two existing datasets: GDELT and w2e, specifically for evaluating event detection and tracking approaches. We show the efficacy of our method by evaluating it on the ground-truth chains.  We leave for future work the improvement in recall by clustering news articles through incorporation of more robust text representations like Doc2Vec. RevDet can also be extended easily to work for streaming news data and this can lead to a truly automated and robust event classifier and an event search engine.

	\bibliographystyle{ACM-Reference-Format}
	\bibliography{sample-base}

\end{document}